\documentclass[a4paper,USenglish]{dagrep-v2018}

\usepackage[utf8]{inputenc}
\usepackage{microtype}
\usepackage{csquotes}
\usepackage{booktabs}
\usepackage{pifont}

\newcommand{\ahref}[2]{\href{#1}{\nolinkurl{#2}}}

\subject{Report from Dagstuhl Seminar 18122}
\title{Automatic Quality Assurance and Release}
\titlerunning{18122 -- Automatic Quality Assurance and Release}

\author[1]{Bram Adams}
\author[2]{Benoit Baudry}
\author[3]{Sigrid Eldh}
\author[4]{Andy Zaidman}
\author[5]{Gerald Schermann}
\authorrunning{Bram Adams, Benoit Baudry, Sigrid Eldh, Andy Zaidman, and Gerald Schermann}
\affil[1]{Polytechnique Montreal, CA, \texttt{bram.adams@polymtl.ca}}
\affil[2]{KTH Royal Institute of Technology - Stockholm, SE, \texttt{baudry@kth.se}}
\affil[3]{Ericsson AB - Stockholm, SE, \texttt{sigrid.eldh@ericsson.com}}
\affil[4]{TU Delft, NL, \texttt{a.e.zaidman@tudelft.nl}}
\affil[5]{University of Zurich, CH, \texttt{schermann@ifi.uzh.ch}}

\seminarnumber{18122}
\semdata{March 18--21, 2018 -- \href{http://www.dagstuhl.de/18122}{http://www.dagstuhl.de/18122}}

\volumeinfo
  {Bram Adams, Benoit Baudry, Sigrid Eldh, Andy Zaidman, and Gerald Schermann}
  {5}
  {Automatic Quality Assurance and Release}
  {8}
  {03}
  {1}
\DOI{10.4230/DagRep.8.3.1}

\subjclass{
,
,
,
,
,
}

\begin{document}

\maketitle

\begin{abstract}
This report documents the program and the outcomes of Dagstuhl Seminar 18122 ``Automatic Quality Assurance and Release''. The main goal of this seminar was to bridge the knowledge divide on how researchers and industry professionals reason about and implement DevOps for automatic quality assurance. Through the seminar, we have built up a common understanding of DevOps tools and practices, but we have also identified major academic and educational challenges for this field of research. 

\end{abstract}


\section{Executive Summary}

\summaryauthor[Bram Adams, Benoit Baudry, Sigrid Eldh, and Andy Zaidman]{Bram Adams (Polytechnique Montreal, CA)\\Benoit Baudry (KTH Royal Institute of Technology - Stockholm, SE)\\Sigrid Eldh (Ericsson AB - Stockholm, SE)\\Andy Zaidman (TU Delft, NL)}
\license

The seminar explored the relationship between DevOps and quality assurance from a software engineering perspective. DevOps has been gaining traction since around 2012, with initiatives formed both in industry and academia. While the importance of DevOps as an enabler in higher quality software is intuitively clear to both industry and academia, we have discussed commonalities in views, but also the challenges that lie ahead for this discipline. 

In essence, human factors are very important, because DevOps is not only a technology, it is a way of working and organizing teams. In this light, we have also discussed the resistance that some team members or even entire organisations seem to have towards automating quality assurance through DevOps. Section~\ref{sec:human_factors_devops} summarizes a group discussion that eventually triggered a set of reflections on this topic of human aspects of DevOps. Yet, we have also discussed how DevOps can be an enabler for onboarding new team members through the availability of a standardized DevOps infrastructure (Section~\ref{sec:onboarding}). The whole group observed the general lack of empirical evidence on the importance and benefits of DevOps in modern software engineering. This final point is tightly connected to another important theme in our discussion: educating software engineers in the ways and associated technologies of DevOps.

The main goal of this seminar was to bridge the knowledge divide on how researchers and industry professionals reason about and implement DevOps for automatic quality assurance. Through the seminar, we have built up a common understanding of DevOps tools and practices, but we have also identified major challenges for this field of research as well as for the teaching of DevOps principles and practices. 

This Dagstuhl was a 2.5 day seminar, which we structured around 4 invited talks that served as keynotes to introduce key topics for discussions. These talks, summarized in Sections~\ref{sec:talk_lwakatare} through~\ref{sec:talk_eldh}, were given at the beginning of each morning and afternoon to inspire topics for further discussions on a given topic. The group split into smaller sub-groups after each keynote, in order to focus discussions and reflections on a specific topic. All these discussions have been summarized in the form of a blog post, while in Dagstuhl, and are provided in this report. 

In addition to keynotes and subgroup discussions, we had a plenary session to start the seminar, where each participant had 2 slides for a short introduction; we had a ``speed-dating'' session on Tuesday evening; and we organized a panel discussion about the future of the field on the last morning (Section~\ref{sec:panel_next}). 


\tableofcontents


\section{Overview of Talks}
\label{sec:talks}

\abstracttitle{Understanding the DevOps Concept}
\label{sec:talk_lwakatare}
\abstractauthor[Lucy Ellen Lwakatare]{Lucy Ellen Lwakatare (University of Oulu, FI)}
\license

The DevOps concept in the software industry was first coined in 2009. A decade later, its actual meaning, characteristics and impacts are starting to get established in the software industry and in academia. The key idea of DevOps is the automation of activities that span software development and operations, in order to rapidly and reliably release software changes to the target environment. This requires a change of mindset and consideration of other non-technical aspects, especially when taken into context inside an organization. However, the practices promoted by DevOps do not ignore 
prior research, since DevOps is “standing on the shoulders of giants” in the form of Agile and Lean practices.


\abstracttitle{Perspectives on Teaching a DevOps Course}
\label{sec:talk_parnin}
\abstractauthor[Christopher J. Parnin]{Christopher J. Parnin (North Carolina State University - Raleigh, US)}
\license

Continuous deployment is a software engineering process where incremental software changes are automatically tested and frequently deployed to production environments. Unfortunately, the skills required in automating and accelerating changes to production infrastructure require expertise and training that is even more rare and highly sought than data science skills. Parnin has been teaching a DevOps course since 2015. In the course, students learn how to build an automated deployment pipeline and manage the infrastructure required by a program.

To support the development of the concepts and practices related to this course, Parnin co-organized  continuous deployment summits with companies such as Google, Facebook, and Netflix to understand how industry is using these tools and processes in practice. To support learning, Parnin is using active learning and mutual engagement through the use of workshops, which enable students to work hands-on with tools and code examples. Students primarily work on course projects with four milestone deliverables: configuration management + build; test + analysis; deployment + infrastructure; monitoring + operation. An example task for a milestone would be: provisioning and configuring a Jenkins server automatically using Ansible.

Teaching challenges include costs and risks associated with operating infrastructure, students left behind because of skill gaps, and difficulty of assessing the quality of infrastructure for grading and evaluation. Curriculum challenges include keeping practices up to date, deciding on the balance between teaching tools vs. ideas and the need to validate the processes and practices with empirical studies.


\abstracttitle{Agile Transformation Journey to DevOps from a Test Quality Perspective at Ericsson}
\label{sec:talk_eldh}
\abstractauthor[Sigrid Eldh]{Sigrid Eldh (Ericsson AB - Stockholm, SE)}
\license

The Agile process paradigm changed the way industries work, despite the lack of scientific basis. Agile ways of working promised fast deliveries and high changeability. During Ericsson’s transformation of its Ultra-Large-Scale Systems, a key focus was to establish and evaluate what was lost and gained in this fundamental shift. Test automation improvements are in the center of this change supported by the continuous integration practices. Closing the gap to operations by improving upon the release procedures is still complicated, since it is aiming for manual steps to be fully automated. As Ericsson's vast experience on non-functional tests and systems view are key practices, e.g., robustness testing, one can still find difficulties in fully automating these abilities, as they heavily depend on the hardware, i.e., test environment. Complete system ``end-to-end'' practices are key elements of quality assurance. The goal to test with operation data used directly to achieve a more realistic execution profile, is performed through capturing data with real-time analytics and e.g. utilizing data from various logs.  The early use of actual data for test, will change the way testing is performed, and will be a grand shift enabling new opportunities to assess quality.


\abstracttitle{Challenges of Automatic QA and Release in Industry}
\label{sec:talk_wright}
\abstractauthor[Hyrum K. Wright]{Hyrum K. Wright (Duolingo - Pittsburgh, US)}
\license

Drawing on years of experience in open source, large companies, and startups, Hyrum described the two major factors involved in implementing good QA and release processes, as well as a challenge for those present to impose the state of the art into practice. The two areas of focus are the social and the technical issues involving solid QA and release processes, with the social issues being the most difficult to overcome. Our challenge as a community is to build technology and standards that enable social improvement in areas such as testing and automation. Time will tell if we can rise to the challenge.



\section{Working Groups}
\label{sec:working_groups}

The topic of quality assurance and release covers multiple areas.
It includes not only DevOps-concepts from a technical perspective (e.g., build, test, deployment automation), but also human factors as it changes how teams interact with each other.
Furthermore, in the breakout groups we discussed differences between industry and open source projects (e.g., the process of ``onboarding'' new employees or contributors to a project), how we can transfer verification or testing techniques from software engineering to the DevOps world (e.g., applying those concepts on the pipeline-level), and how bots can support developers and release engineers.

\abstracttitle{Desirable Properties of Pipelines and How to Verify Them}
\abstractauthor[Bram Adams, Foutse Khomh, Philipp Leitner, Shane McIntosh, Sarah Nadi, Andrew Neitsch, Christopher J. Parnin, Gerald Schermann, Weiyi (Ian) Shang, and Hui Song]{Bram Adams (Polytechnique Montreal, CA), Foutse Khomh (Polytechnique Montreal, CA), Philipp Leitner (Chalmers University of Technology - Göteborg, SE), Shane McIntosh (McGill University - Montreal, CA), Sarah Nadi (University of Alberta - Edmonton, CA), Andrew Neitsch (Cisco Systems Canada Co. - Toronto, CA), Christopher J. Parnin (North Carolina State University - Raleigh, US), Gerald Schermann (University of Zurich, CH), Weiyi (Ian) Shang (Concordia University - Montreal, CA), and Hui Song (SINTEF ICT - Oslo, NO) }
\license

\begin{figure}[h]
	\centering
	\includegraphics[width=0.85\columnwidth]{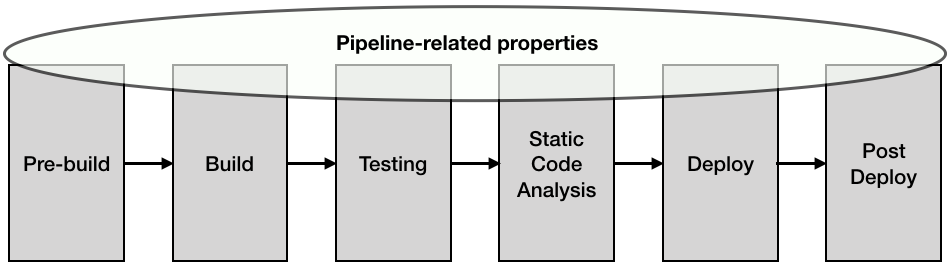}
	\caption{Various stages of a pipeline}
	\label{fig:pipeline}
\end{figure}

Release/deployment/delivery pipelines, what we call ``pipelines'', are becoming fundamental artifacts in modern software organizations. 
These pipelines automate different stages of a deployment process and have the ability to filter out poor quality commits and target specific delivery endpoints (such as a staging or production environment). 
Figure~\ref{fig:pipeline} shows an overview of such a pipeline. 
Broadly speaking, a pipeline includes:

\begin{enumerate}
	\item \emph{Pre-build}: Making decisions about which configurations/platforms/variants to check throughout the pipeline.
	\item \emph{Build}: Transforming the sources into deliverable format.
	\item \emph{Testing}: Execution of various types of automated tests (e.g., unit, integration, performance, regression).
	\item \emph{Static Code Analysis}: Automated analysis of source code for common problematic patterns.
	\item \emph{Deploy}: Shipping newly built deliverables to staging and production environments.
	\item \emph{Post Deploy}: Making sure that the released application performs/behaves as expected (e.g., through monitoring and production testing).
\end{enumerate}

The recent shift towards continuous deployment has resulted in the need to deploy changes into production environments at an ever faster pace. 
With continuous deployment, the elapsed time for a change made by a developer to reach a customer can now be measured in days or even hours. 
Such ultra-fast and automatic changes to production means that testing and verifying the design and implementation of the pipelines is increasingly important. 
The high-level idea is that pipeline artifacts are also susceptible to the same sorts of problems that we encounter when writing code (e.g., defects, anti-patterns); however, pipeline quality assurance practices are rarely applied (if at all). 

So how can we apply quality assurance practices to pipelines? 
In this blog post, we begin by defining properties that stakeholders would consider desirable (or undesirable) in the pipeline and each phase within it. 
We provide examples of what could go wrong if a property is violated and some potential avenues for verifying these properties in practice. 
Finally, future research and tool smiths can leverage these properties in the design of better pipeline engines and tools.

These observations are based on brief discussions between DevOps researchers and practitioners. 
Our goal is not to advocate for the completeness of this list, but rather to start a discussion within the community around these properties.

\subsubsection{Overarching Pipeline Properties}
\newcommand*\rot{\rotatebox{90}}
\newcommand*\OK{\ding{51}}

\newcolumntype{P}[1]{>{\centering\arraybackslash}p{#1}}

\begin{table}[h!]
	\centering
	\begin{tabular}{@{}lP{0.5cm}P{0.5cm}P{0.5cm}P{0.5cm}P{0.5cm}P{0.5cm}P{0.5cm}@{}}
		\toprule
		&  \rot{Pipeline} & \rot{Pre-build} & \rot{Build} & \rot{Testing} & \rot{\shortstack[l]{Static Code\\Analysis}} & \rot{Deploy} & \rot{Post Deploy}  \\ 
		\midrule
		Scalability                  & \OK &     &     & \OK &     &     &     \\ 
		Repeatability                & \OK &     & \OK & \OK &     & \OK & \OK \\ 
		Simplicity/Understandability & \OK &     &     &     &     &     &     \\ 
		Trustworthiness              & \OK &     &     &     &     &     &     \\ 
		Security                     & \OK &     &     &     &     &     &     \\ 
		Robustness                   & \OK &     &     &     &     &     &     \\ 
		Availability                 & \OK &     &     &     &     & \OK & \OK \\ 
		Velocity/Speed               & \OK &     & \OK & \OK & \OK & \OK &     \\ 
		Cost                         & \OK &     &     &     &     &     &     \\ 
		Coverage                     &     & \OK &     & \OK &     &     &     \\ 
		Incrementality               &     &     & \OK &     & \OK & \OK &     \\ 
		Correctness                  &     &     & \OK &     & \OK & \OK &     \\ 
		Regularity                   &     &     &     &     &     &     & \OK \\ 
		Traceability                 & \OK &     &     &     &     &     & \OK \\ 
		Independence                 &     &     & \OK &     &     &     &     \\ 
		\bottomrule 
	\end{tabular} 
	\caption{Overview of pipeline properties and in which phases they should be enforced}
	\label{tab:pipeline_properties}
\end{table}

\begin{itemize}
	\item \emph{Scalability/Cost}: A pipeline needs to be able to deal with a high load of commits coming in
	\item \emph{Repeatability}: Treating a pipeline as an artifact itself, installing the same pipeline under the same configuration should lead to the exact same results/outcome when fed with the same data 
	\item \emph{Simplicity}: It should be easy to understand of what phases/stages a pipeline consists of and how changes flow through the pipeline.
	\item \emph{Trustworthiness}: Trusting tools or bots that have privileges to modify the configuration or properties of the pipeline.
	\item \emph{Security}: Making sure that all phases/steps along the pipeline are reasonably secured (e.g., ports only open when needed)
	\item \emph{Robustness/Availability}: Involves the different phases/steps that need to be up and running (e.g., staging environments).
	\item \emph{Velocity/Speed}: The time needed to run through all the phases of the pipeline.
	\item \emph{Traceability}: At which specific stage/phase of a pipeline is a (code) change currently? Is the pipeline able to route each change along the right sequence of tasks and to detect each problem as soon as possible? 
\end{itemize}

\subsubsection{Do I really need to care about the pipeline?}

At this point, you may be wondering how violations of the above properties manifest themselves as problems in practice. 
My code works well, do I really need to care about the pipeline and this long list of properties you provide? 
The answer is yes! 
Before diving into the details of each of the above properties and how they are related to each stage of the pipeline, let us talk about some real-world examples first. 

Starting with the importance of the scalability of the pipeline: Major projects like Openstack receive so many commits per hour and have thousands of tests to be executed, such that they are not able to trigger the full CI build for each individual commit. 
Instead, OpenStack’s testing automation team\footnote{\ahref{http://web.archive.org/web/20180608001910/https://archive.fosdem.org/2014/schedule/event/openstack_testing_automation/}{https://archive.fosdem.org/2014/schedule/event/openstack_testing_automation/}} came up with custom algorithms to run the CI build on a group of commits, while not losing the ability of singling out the specific commit responsible for failing the (group) build.

Another property to think about is security. 
If my product code is secure, as well as that of the data servers it connects to etc., what can possibly go wrong? 
Well, the server that you eventually deploy to, e.g., for others to download your application from, may itself be insecure. 
A recent example is Eltima’s Media Player\footnote{\ahref{http://web.archive.org/web/20180608001909/https://www.macrumors.com/2017/10/20/eltima-software-infected-with-malware/}{https://www.macrumors.com/2017/10/20/eltima-software-infected-with-malware/}}, where a trojan was inserted in the Mac app after it was already deployed to the download server. 
An older example is a backdoor code change that was added into the Linux kernel CVS mirror of the official BitKeeper repo\footnote{\ahref{http://web.archive.org/web/20180608001909/https://freedom-to-tinker.com/2013/10/09/the-linux-backdoor-attempt-of-2003/}{https://freedom-to-tinker.com/2013/10/09/the-linux-backdoor-attempt-of-2003/}}, without any trace where it came from (i.e., it seemingly escaped code review, and likely was hacked into the repository). 
Luckily, it did not end up in kernel releases, since the commit was just inserted into a mirror, but who knows what would have happened if this was not the mirror server!

\subsubsection{What are examples of what I can do to avoid this?}

In terms of scalability, one solution is that instead of letting pipelines process and/or run every type of file (e.g., shell script) and third-party library as part of a code change that is checked in, the pipeline should verify the type of changes that are allowed to trigger the pipeline, as well as specify explicitly who is allowed to check in what (humans, bots, ...).

To ensure that all servers you use in your pipeline have the same properties (e.g., security, installed software etc.), we can use the concept of Pipeline-as-Code (e.g., \ahref{http://web.archive.org/web/20180608001909/https://jenkins.io/blog/2017/09/25/declarative-1/}{https://jenkins.io/blog/2017/09/25/declarative-1/}), which allows specifying the pipeline infrastructure in textual form and to automatically instantiate the pipeline, similar to infrastructure-as-code. 
This enables instantiating identical pipeline instances across different machines, to track changes in the specification (Git), etc. 
Another option is to implement the concept of \emph{flow testing}. 
\emph{Flow testing} can be performed in the style of integration testing by validating that a given test commits indeed performs as expected. 
Did the pipeline hit the major pipeline steps that it should hit (build, test, deployment etc.). 
For a commit that causes a performance problem, was this problem caught at the right stage and flagged? 
Each kind of problem requires a ``barrage'' point behind which we do not want the problem to pass.
Finally, specific steps can be taken to ensure the security of all pipeline servers by using checksums and reproducible builds (i.e., a build process that always generates the same binary, bitwise, for a given input), which can further avoid intermediate tampering with build artifacts.

Now that we have given you a big picture of things, let’s get down to the details a bit. 
Table~\ref{tab:pipeline_properties} summarizes the different stages of the pipeline and the properties we believe each stage should enforce. 
In the following sections, we discuss these properties in more detail and provide examples of what can be done to enforce these properties to avoid problems such as those previously described.

\paragraph*{Pre-build}

Desirable properties: 
\begin{itemize}
	\item \emph{Cost}: The minimal set of configurations/variants/branches should be used within the subsequent phases (Tradeoff with Coverage)
	\item \emph{Coverage}: The largest set of configurations/variants/branches should be used within subsequent phases (Tradeoff with Cost)
	\item \emph{Understandability}: How could the engineers understand the configurations as specified in the pipeline
\end{itemize}

\noindent \emph{Example problem}: It is very easy to waste resources on running test cases on multiple configurations. \\
~\\
\noindent \emph{Example verification}: Through different traceability links, the pipeline can assess if a test case touches certain configuration points and plan accordingly by reducing the number of representative configurations.

\paragraph*{Build}

Desirable properties:
\begin{itemize}
	\item \emph{Correctness}: Dependencies are fully expressed.
	\item \emph{Velocity/Speed}: Builds are completed in a reasonable amount of time (reasonable varies from project to project).
	\item \emph{Incrementality}:  Builds (re)execute the minimal subset of necessary commands to update deliverables without missing any necessary commands.
	\item \emph{Repeatability}: Given the same input, builds perform the same commands (i.e., deterministic builds). Moreover, it should be possible to reproduce a past build in the future (e.g., if a package service goes down or is no longer available).
	\item \emph{Independence}: Builds should be isolated from each other.
\end{itemize}

\noindent \emph{Example problem}: A build phase that does not have the independence property may suffer from builds that interfere with each other. In turn, the build phase may become non-deterministic. For example, builds that are running in parallel or sequentially may access resources from each others’ environments by mistake. \\
~\\
\noindent \emph{Example verification}: A possible step towards checking this property could be to apply a form of taint analysis, i.e., track all outputs of a build and check who reads those outputs. Taint analysis has been effectively applied to the analysis of the surface area that is exposed to security issues (e.g., SQL injections). The same concepts may apply to the leakage of state within the scope of builds. 

\paragraph*{Testing}

Desirable properties:
\begin{itemize} 
	\item \emph{Scalability/Cost}: The testing stage needs to be ``smart'' about its decisions. Depending on the size of the test suite and types of tests available, not every test needs to be run for each commit. Only tests affected by updated artifacts should be run. Good traceability links between code and the test suites, as well as test prioritization can be used to make the testing stage more scalable.
	\item \emph{Repeatability}: Running the same test suite on the same commit should always produce the same result in terms of passed and failed tests. Flaky tests\footnote{\ahref{http://web.archive.org/web/20180608001909/https://testing.googleblog.com/2016/05/flaky-tests-at-google-and-how-we.html}{https://testing.googleblog.com/2016/05/flaky-tests-at-google-and-how-we.html}} are especially problematic for repeatability. One solution is to identify/flag flaky tests in order to have special handling for them. 
	\item \emph{Velocity/Speed}: Execution time of test suites is a major bottleneck in the overall velocity of the pipeline. In this phase, velocity/speed is related to the scalability/cost property since smarter test selection will probably eventually lead to better speed of the testing phase.
	\item \emph{Coverage}: As many possible variants/configurations of a product need to be tested, without sacrificing speed.
\end{itemize}

\noindent \emph{Example problem}: If no test prioritization/selection strategy is used, large amounts of testing resources can be wasted on not impacted artifacts, delaying the delivery. \\
~\\
\emph{Example verification}: With a predefined set of mappings between code and tests, given the code changes, the pipeline should trigger and only trigger those tests.

\paragraph*{Static Code Analysis}

Desired properties:
\begin{itemize} 
	\item \emph{Incrementality}: Only the needed analysis is applied in the release pipeline. For example, the static analysis is only applied on the code that is impacted by the code change. (may remove: Capture the intended properties of the analysis)
	\item \emph{Correctness}: A static analysis should yield a low rate of false positives, since false positives reduce the trustworthiness of results and lead of gain adoption from the practitioners. 
	\item \emph{Performance}: The static code analysis should be able to finish within a reasonable time, since a long duration of the analysis will affect the deliverable of the product into the next step in the pipeline (e.g., testing or deployment).
\end{itemize}

\noindent \emph{Example problem}:
A typical off-the-shelf static analysis tool often report a large number of issues, while not all of them are of interest to impact the release pipeline. 
Reporting all the issues, or having all the issues to determine the next step in the pipeline is problematic in practice. \\
~\\
\emph{Example verification}:
A dataset with the issues that may be detected by static code analysis needs to be labeled into whether they are of interest of the practitioners to impact the release pipeline. 
To test the correctness of the static analysis, a randomly generated sample from the dataset is the test input of this phase and the precision and recall with threshold can be used to assert whether the output is satisfactory.

\paragraph*{Deploy}

Desired properties: 
\begin{itemize} 
	\item \emph{Repeatability}: The deployment results of the software should not be impacted by the deployment environment.
	\item \emph{Availability}: It is desirable that the service is continuously available during deployment, instead of having to choose between either working nights/weekends to release, or bringing down the service during peak times.
	\item \emph{Velocity}: A regular release rhythm and/or immediate releases that let developers say that their feature is truly ``done done'' since it is released to production, can aid development velocity.
	\item \emph{Incrementality}: An incremental release that initially sends only a fraction of traffic to the new version, and is ramped up until the new version is handling all traffic, can limit the ``blast radius'' of a problematic release.
	\item \emph{Correctness}: Deploying software can be risky. A correct process that will not leave the service in a broken state on error is highly desirable.
\end{itemize}

\noindent \emph{Example problem}:
The new version of the service requires configuration for a new feature, but the new configuration has not been applied to the target environment. After deploy, the new version crashes on startup with an error message about invalid configuration. \\
~\\
\emph{Example verification}:
Recently added and changed acceptance tests are run against the newly-deployed service before it is exposed to external clients. 
Automatic rollback is triggered if the service crashes, or the tests fail.

\paragraph*{Post Deploy}

Desirable properties:
\begin{itemize} 
	\item \emph{Availability}: Are the deployed artifacts available (e.g., appropriate ports open, heartbeat)
	\item \emph{Regularity}: Configuration and code changes perform within nominal operational metrics.
	\item \emph{Traceability}: Data is ``collectable'' on a continuous basis from various parts of the system, and configuration changes should be auditable and mapable to code changes.
	\item \emph{Repeatability}: To what extent is the infrastructure resilient to changes in external dependencies, versions, and tools.
\end{itemize}

\noindent \emph{Example problem}:
Inadvertently reusing a feature flag’s name can make dead code active again, as happened with the Knight Capital bankruptcy. \\
~\\
\emph{Example verification}:
Turning on an old feature flag could violate two properties: a) \emph{Traceability}, code associated with a feature flag may not have been recently changed, which could set off a warning. b) \emph{Regularity}, the performance of the code with the wrong feature flag may generate metrics that are not consistent with recent range of metrics.


\abstracttitle{Human Factors in DevOps}
\label{sec:human_factors_devops}
\abstractauthor[Lucy Ellen Lwakatare, Tamara Dumic, Sigrid Eldh, Daniele Gagliardi, Andy Zaidman, and Fiorella Zampetti]{Lucy Ellen Lwakatare (University of Oulu, FI), Tamara Dumic (Ericsson Research - Stockholm, SE), Sigrid Eldh (Ericsson AB - Stockholm, SE), Daniele Gagliardi (Engineering Ingegneria Informatica S.p.A - Roma, IT), Andy Zaidman (TU Delft, NL), and Fiorella Zampetti (University of Sannio - Benevento, IT) }
\license

\subsubsection{Introduction}

Change of process, ownership and technology has a direct impact on individuals, organizations and groups of people. DevOps is a process that includes philosophy, mindset changes, and also changes in human interactions and communications. 
Tackling human factors is beyond the mere technology shift that DevOps introduces. 
One can say that transforming to DevOps is a software process change that embodies the human performing the process as an essential part. Specifically, DevOps adoption implies a transformation of the entire software community including the way to release, monitor, and interact with the users of the system. 
Through its extensive approach to human tasks automation, it also transforms ownership, organization and could be considered a major technology change in the way we create, produce and use the software. 
As a process, it enables new technologies and makes data collection shared and available for analysis in a completely new way. 
The following items summarize the five main challenges that we have identified as key topics belonging to the human factors of process change:

\begin{enumerate}
	\item It is hard to convince people to do/adopt DevOps
	\item Need to have a shared approach/vision of all stakeholders/organizations involved  
	\item Difficulty in accountability/taking ownership for people
	\item Overcoming fear of new technology and change
	\item Other topics
\end{enumerate}

For each of these five challenges, we have clarified the context and given examples in order to avoid ``misunderstandings''. 
Moreover, when possible, we have highlighted some possible solutions in addressing them. 

We have identified and observed a number of signs of difficulties in the DevOps transformation, e.g., resisting change, split responsibility, ownership, and assumptions and expectations on the software process and life-cycle that might not be accurate, as well as basic lack of skills in different fields to enable the process change. 
This list can be expanded and discussed, to aid others in the DevOps change.

Moreover, it is important to highlight that, in our discussion, we have also identified some key human factors that go beyond the scope of this challenge that could be addressed by expertise in other scientific communities, enabling cross-science collaborations.

\subsubsection{Challenges}

\paragraph*{It is hard to convince people to do DevOps }

DevOps transformation and adoption in an industrial setting, including large organizations and those in public sector, requires change agents to make a convincing argument ``business case'' for its enactment. 
The lack of empirical analysis aimed at demonstrating the advantages of using a DevOps approach despite the initial effort required and the lack of contextualization inevitably implies resistance to change and, consequently, resistance to the adoption of new technologies and/or practices. Further, not having scientific evidence of the advantages of adopting a DevOps process in different contexts/domains, but having only anecdotes, makes it difficult to create a convincing argument to be used with the various stakeholders involved in the change. 
As a matter of fact, automation in DevOps not only requires new resources but also changes existing processes and organizational structures thereby affecting people’s work and competences. 

It is not unknown that DevOps emerged as a hype word in the software industry in the past decade. 
Leading software companies, such as Facebook, Google and Microsoft, reported a paradigm shift towards multiple release of new features to end users on a daily basis instead of the typical lengthy release cycle times. 
While technical aspects, and the ideas behind them, are relatively clear, onboarding them in practice requires the acknowledgment by the people involved in it. 
The latter is particularly true since often people either out of fear or out of laziness tend to reject new solutions when they have something equivalent and working at their disposal (solutions perceived to be sufficient). 
A different aspect to take into account is related to the initial investment (effort) required in order to setting up a Continuous Deployment pipeline. 
In this context it is very important and needed to make clear the return value (ROI) and the impacts for the different stakeholders involved (e.g., it improves release cycle times, early identification of problems and reduction in the overall costs of maintenance activities). 
This is not so easy to achieve since that in many cases the change process includes different stakeholders having different backgrounds and needs. 

A way forward would be to ensure that people are convinced that the new practices are worth the effort and investment.
In order to obtain the above goal it is very important and needed a strong collaboration between both industry and academia in performing studies that are able to highlight its benefits. 
Indeed, it is well known that scientific facts are more easy to be believed compared to anecdotal evidence.
The lack of scientific evidence has led industries to look at DevOps and continuous deployment in general as ``black magic''.

Another key aspect that needs to be taken into account is related to the awareness and visibility of the whole ``systems thinking'' rather than isolation of system life cycle phases. 
Very often developers do not know what is the effect of their ``little'' contribution on the whole system under development. 
A common statement to avoid/remove is: ``But it works/worked on my machine''. 
The latter can be obtained augmenting the developers' knowledge. 
It is very important that each developer knows that their contribution/change can break the whole system when integrating it and this will inevitably imply that other developers/teams can be blocked as a consequence of their failures. 
Also in this case, scientific evidence can help in augmenting the awareness of developers working in a large group or in a large project on the fact that they can block the whole system even with a little change. 
Finally, another interesting point to highlight is related to the economic perspective in the sense that it can be interesting for industries to know the economic loss in missing developers' contribution awareness. 

\paragraph*{Need to have a shared/approach vision of all stakeholders/organizations involved}

Implementing DevOps is not something that one single team or even one single person can decide. 
It requires that the development team, the operational team, and the customer are willing to do so. 
As an example, if the development team wants to do it, but the company is unwilling to pay for the operational/monitoring aspect, you cannot move to DevOps.

Another aspect to consider is the team composition. 
Indeed, many teams also include IT consultants but, if many consultancy firms are cooperating, it is required to have an agreement amongst them.

We have the feeling that the customers are the driving force behind moving to DevOps and not the service company (or development team). 
It is the customer that approaches them with a buzzword like DevOps and asks for it. 
There are also cases in which the customers want and pay for a product, without explicitly making mention of the quality assurance practices/standards that need to be upheld. 
This sometimes leaves the door open for service companies to cut back on quality assurance during the engineering/development phase, only to get a maintenance contract for the product that they are delivering afterwards. 
In other words, sometimes the customers decide to not pay for testing (and also quality checking of the software product) since their main aim is to have a prototype ready for usage. 
Only in a subsequent step, they will define a new contract in order to check whether the software product effectively fits their needs. 
As a Software Engineering community, we know that doing this will inevitably increase the cost belonging to finding defects but, most important, the cost of fixing them. 
DevOps adoption can reduce this cost since each developer can immediately see whether her change is inducing a failure (test case failure, not respect of rules/standards to follow in terms of quality, etc) and can easily identify where the failure is located inside her code. 

As a consequence, it is very important to have a shared/common approach/vision between different stakeholders/organizations involved. 
The main problem to address is that, very often, different stakeholders show different backgrounds and, most important, have different goals. 
Of course, this is not so easy to obtain when each organization is close-minded in the sense that it does not want to look at the whole picture, but is only focused on its own needs. 
Finally, it is very important that in order to be competitive on the market you cannot focus only on the cost but you have to focus on the quality of what you are going to put on the market. 

\paragraph*{Difficulty in accountability/taking ownership for people}

As already highlighted, DevOps adoption means automation but also collaboration between people. 
Obviously, collaboration implies that each party involved shall take the ownership for something inside the collaborative process but also for the relations it has with other actors. 
A ``silo mentality'' is what we actually have in different teams involved in the realization of the same software product. 
In other words, a team or department shares common tasks but derives their power and status from their group with no accountability about relations. ``I do my best in my tasks. Period''. 
An example can be the one related to a ``Misunderstanding of requirements'': when the developer appears to be responsible for a requirement not well-understood, she usually says ``the functional analyst gave me wrong information''. 
And the analyst will blame someone else for something else. 
This is related to the fact that often companies or, maybe better, single managers act following a ``blame culture'': if something goes wrong, the first action is finding the culprit, not the solution. 
The latter is part of the corporate culture.

A first step towards the promotion of accountability could be to define as much as possible what each actor is accountable for, and how what she does will impact the whole process: knowing the ``what'' is necessary to find the ``how'' (how can I improve to serve better the others? Do I see improvements that others can do to help me to improve?). 
As an example, in many situations it is not taken for granted what is Dev accountability and Ops accountability. 
More in general, Dev accountability is related to verification and validation of the compliance between software requirements and the product under development. 
On the other hand, Ops accountability is related to availability and reliability of software services in production systems. 
If we introduce explicitly the concept of ``accountability for quality'', it becomes easy to see that each actor cannot act as she was in an isolated context with no relations with others: well-defined requirements implies well-developed software that requires an effective test suite (set of test cases). 
If test cases are considered as ``executable specifications'', they can help to define in a better way the requirements. 
As also suggested in the other two challenges already exploited, in developing a software product you cannot have tasks/phases considered as second-class citizens (unfortunately, this usually happens for test cases that are not written before starting the real development but also not written together with production code).  

Another aspect is that acting within DevOps processes (and ultimately being accountable for something) means also to commit on continuous learning (not just new tools or new processes, but also gaining awareness of the big picture). 
Continuous learning can be hard to accomplish (people need time, resources). 
Another obstacle in being accountable for something is related to the fear of people in being responsible for something that goes wrong (different from what is desired). 
Related to the continuous learning it is important to consider non-technical stakeholders as functional analysts and/or clients. 
Since that, very often a company has to see what the clients want as a cornerstone: ``the customer is always right'' causes people have unsustainable pressure, and asking people for being accountable for something cannot be required.

\paragraph*{Overcoming fear of new technology and change }

A change is often associated with uncertainty and fear, especially if it has direct impact on individuals. 
Automation of different process steps is not an exception to that. 
It can lead to showing resistance and loss of motivation for some individuals, especially in cases when an existing expertise is threatened by the automation. 
This is typically a case when an individual has been doing a type of work that can be replaced by an automated process (for example a tool or a script). 
If at the same time a threshold for learning new technologies is simply too high for that individual, the resistance and loss of motivation tend to be even greater. 
Driving change is certainly not a trivial thing to do, especially in large organizations. 
It requires a lot of communication and having a dialogue with those individuals who might have concerns related to the change. 
Creating an environment where each individual feels included and listened to, improves chances for getting a buy-in. 
Also, in the process of communicating a change it is very important to keep coming back to the big picture (the reason for the change) and the benefits that the change will bring.

\paragraph*{Other challenges}

This last category is a ``remaining'' section of areas that could be observable issues related to human factors, and not necessarily fitting in the above sections. 
We will here shortly explain each of these challenges:

\emph{Awareness about what quality really is:}
A common theme is that people often tend to underestimate the work that is happening, especially in the context of test and quality assurance. 
If testing practices are not a real part of the software development, there is bound to be resistance to, e.g., automation of tests -- since the people are not mature in testing -- and cannot understand the long term benefits of such investments. 
There is not sufficient nuance in categorizing phases of company growth that can be used -- as maturity on collaborations and understanding of both development, operations technical aspects of CI/CD is still flawed.

\emph{Organization set-ups:}
Since some organizations are set up in ways where often development and operations have separated -- the ownership, and the challenge DevOps is the management of both of them -- this is must engage both aspects. 
For some communities, this poses a huge difficulty and prevention of progress.
There is not sufficient studies on solving this -- and it might even not be in peoples own interest to, e.g., loose their ``power'' over a part in the process -- when new ways are asking for shared ownership.

\emph{Education and fact-based benefits of doing things:}
There are two aspects here. 
One is that education should be based on scientific evidence -- and DevOps is also addressing operations, release aspects, monitoring aspects etc. Second, gathering information and research from these areas.
A special break out report on this from Dagstuhl, named ``Dagstuhl - DevOps and the Need for Improved Evidence from Industrial Studies'' dives deeper into this topic (see Section~\ref{sec:evidence_industrial_studies}). 

\subsubsection{Solutions and Considerations}
\begin{itemize}
	\item Knowledge/education can overcome  Challenge 1,2,4
	\item Scientific studies (proof of benefit) can overcome 2
	\item Trade-off between ``ok to break'' (and being fast and have resilient software) vs the culture of ``do not break the builds''... 
	\item Promoting the value of continuous learning can address 3: being aware of the relations and the impact requires training, going towards the concept of cross-functional teams. It is not needed that a developer becomes an expert sysadmin, but it's crucial that she knows how the quality of her code (as an example) impact on operations, and ultimately on software quality attributes (security, performance, availability, reliability)
	\item Blameless build features -- ``it is not the person, it is the process''... what process failure led to this happening...
	\item Despite good processes in place, humans can fail (tiredness, psychological personal situation, etc): we should accept it ``by culture''. But maybe good processes could prevent consequences of human failures (i.e., checklists to face the emergency, etc)
	\item Can we provide tools to mitigate human issues?
\end{itemize} 


\abstracttitle{Bots for DevOps}
\abstractauthor[Martin Monperrus, Benoit Baudry, Moritz Beller, Benjamin Danglot, Zhen Ming (Jack) Jiang, Vincent Massol, Oscar Luis Vera Perez, and Hyrum K. Wright]{Martin Monperrus (KTH  Royal Institute of Technology - Stockholm, SE), Benoit Baudry (KTH Royal Institute of Technology - Stockholm, SE), Moritz Beller (TU Delft, NL), Benjamin Danglot (INRIA Lille, FR), Zhen Ming (Jack) Jiang (York University - Toronto, CA), Vincent Massol (XWIKI - Paris, FR), Oscar Luis Vera Perez (INRIA - Rennes, FR), and Hyrum K. Wright (Duolingo - Pittsburgh, US) }
\license

Parts of the automation in the DevOps pipeline are provided by so-called ``bots''.
Who are they? What makes them ``bots''? 
In this post, we explore the essence of bots for DevOps. \\
~\\
We first consider some concrete examples of bots in the context of DevOps. 
Then we derive a set of common characteristics from these bots. 
Finally, we discuss the values of these bots and various challenges associated with developing bots for DevOps.

\paragraph*{Examples:}
Let us first consider a list of concrete examples of bots (meant to be illustrative, not comprehensive): \\
\noindent A bot that ...
\begin{itemize}
	\item replaces all calls to deprecated methods with the new methods (\emph{DEPRECATEBOT})
	\item checks licensing and intellectual property (\emph{CLABOT})
	\item spells check comments and identifiers (\emph{NAMEBOT})
	\item reformats code according to some coding styles/convention (STYLEBOT)
	\item applies static analysis to detect bugs (\emph{BUGBOT})
	\item applies program repair techniques to fix bugs (\emph{REPAIRBOT})
	\item says ``Welcome'' to new contributors when they post their first pull-request (\emph{WELCOMEBOT})
\end{itemize}

\begin{figure}[h]
	\centering
	\includegraphics[width=0.75\columnwidth]{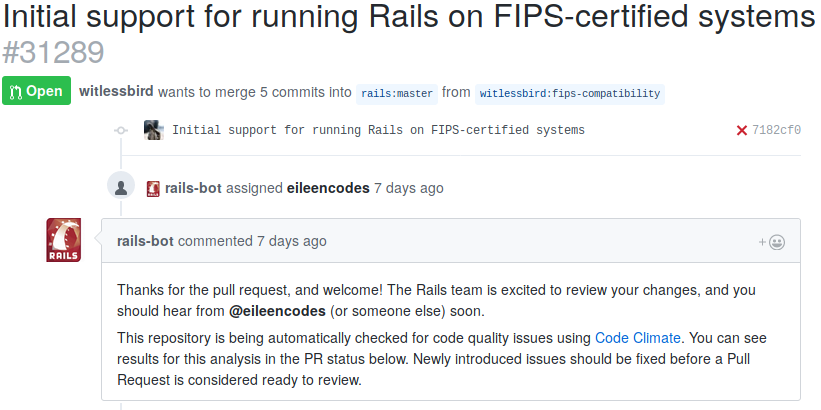}
	\caption{A welcome bot giving a greeting to a newcomer to the project. (Source: M. Beller, An Empirical Evaluation Of Feedback-Driven Development, 2018)}
	\label{fig:welcome_bot}
\end{figure}

\noindent Those bots have very different characteristics, but what is sure is that: 
\begin{displayquote}
``A bot is not defined by the complexity / size of the task that is automated''
\end{displayquote}

For instance, a WELCOMEBOT like the ``rails-bot'' in Figure~\ref{fig:welcome_bot} can be technically simple, but a REPAIRBOT can be very complex to develop and maintain. 
It seems that what makes a bot is more related to the intention behind the bot. 
This makes us propose not a single definition, but a set of characteristics.

\subsubsection{Characteristics of DevOps Bots}
To us, a bot satisfies at least one of the following definitions (logical ``or'', not ``and''): \\
\noindent A bot ...
\begin{itemize}
	\item does its job in an autonomous manner, not when humans ask for it (a.k.a., it is not a command)
	\item is something that is able to perform transformative actions on software artefacts, including code
	\item performs tasks that could not possibly be done by humans
	\item performs tasks that are traditionally done by humans
	\item outputs something that could also be produced by humans (a.k.a., producing contents which look like human-created content)
	\item is something that interacts with humans with the same medium as humans (e.g., in the pull-request conversation thread)
	\item it's output consumes a human's brain bandwidth
\end{itemize}

Many of those facets focus on the interaction with humans.
To this extent, bots may be most valuable when they need some input by humans to perform a task (otherwise, it can be fully automated, e.g., by the compiler).

\subsubsection{Values of Bots}

Bots can add value to the DevOps pipeline in different ways: 
\begin{itemize}
	\item Bots are required when there is no way to change the root cause of a problem because it is handled by different stakeholders (e.g., you cannot change the tool ``Javadoc'' for handling missing dots at the end of the first sentence, so you create a bot for this).
	\item Second, bots are particularly interesting in the gray area between hard tasks (only done by humans) and tasks that can be completely automated (no bots needed, but an automated tool that does the job).
	\item Third, and quite pragmatically, bots are valuable if they do a task for which one would be ready to pay a human.
\end{itemize}

Even as known bots are deployed in production (e.g., the WELCOMEBOTS from GitHub), we present the following series of grand research challenges to embed bots in DevOps.

\subsubsection{Grand Challenges}
In the following, we define a series of five grand challenges in the area of DevOps.
\begin{itemize}
	\item \textbf{Conversation:}
We do not know how to build conversational code bots, i.e., bots that can propose changes and later respond to the questions of developers about the nature and the rationale of the change, and improve the proposed changes accordingly.

	\item \textbf{Trust:}
	Humans and bots interact. It is teamwork where trust plays an important role. ``How to build and manage bots' reputation?'' ``How do software engineers develop trust in bots?" ``What prevents developers from blindly trusting code bots?" These are open questions for future scientific research. 

	\item \textbf{Interoperability:}
Bot designers have to make assumptions on the software artifacts to be analyzed and transformed. 
How to organize a repo to get the most out of bots? 
Could we set standards for bot-friendly repo organization?

	\item \textbf{Configuration:}
Current bots are mostly hard-wired at deployment time. 
However, rules are not all equally relevant and interesting among many different organizations. 
How to build self-configuring bots that automatically learn the rules, criteria, thresholds to be used?

	\item \textbf{Bot teams:}
In the future, there will be multiple bots working on the same code base: how to set up teams of bots with different goals, characters, which together form a dream bot team?
\end{itemize}

\noindent This makes an exciting research agenda for the DevOps community.


\abstracttitle{Onboarding a Software Project}
\label{sec:onboarding}
\abstractauthor[Vincent Massol, Benoit Baudry, Benjamin Danglot, Daniele Gagliardi, and Hyrum K. Wright]{Vincent Massol (XWIKI - Paris, FR), Benoit Baudry (KTH Royal Institute of Technology - Stockholm, SE), Benjamin Danglot (INRIA Lille, FR), Daniele Gagliardi (Engineering Ingegneria Informatica S.p.A - Roma, IT), and Hyrum K. Wright (Duolingo - Pittsburgh, US) }
\license

When you are developing a project, be it some internal project or some open source project, one key element is how easy it is to onboard new users to your project. 
For open source projects it is essential to attract more contributors and have a lively community. 
For internal projects, it is useful to be able to have new employees or newcomers in general be able to get up to speed rapidly on your project.

\noindent This brainstorming session was about ideas of tools and practices to use to ease onboarding.
Here is the list of ideas we had (in no specific order):

\begin{itemize}
	\item Tag issues in your issue tracker as onboarding issues to make it easy for newcomers to get started on something easy and be in success quickly. This also validates that they're able to use your software.
	\item Have a complete package of your software that can be installed and used as easily as possible. It should just work out of the box without having to perform any configuration or additional steps. A good strategy for applications is to provide a Docker image (or a Virtual Machine) with everything set up.
	\item Similarly, provide a packaged development environment. For example, you can provide a VM with some preinstalled and configured IDE (with plugins installed and configured using the project's rules). One downside of such an approach is the time it takes to download the VM (which could be several GB in size).
	\item A similar and possibly better approach would be to use an online IDE (e.g., Eclipse Che) to provide a complete pre-built dev environment that would not even require any downloads. This provides the fastest dev experience you can get. The downside is that if you need to onboard a potentially large number of developers, you will need some important infra space/CPU on your server(s) hosting the online IDE, for hosting all the dev workspaces. This makes this option difficult to implement for open source projects for example. But it's viable and interesting in a company environment.
	\item Obviously having good documentation is a given. However too many projects still don't provide this or only provide good user documentation but not good developer documentation with project practices not being well documented or only a small portion being documented. Specific ideas:
	\begin{itemize}
		\item Document the code structure
		\item Document the practices for development
		\item Develop a tool that supports newcomers by letting them know when they follow / don't follow the rules
		\item Good documentation shall explicit assumptions (e.g. when you read this piece of documentation, I assume that you know X and Y)
		\item Have a good system to contribute to the documentation of the project (e.g. a wiki)
		\item Different documentation for users and for developers
	\end{itemize}
	\item Have homogeneous practices and tools inside a project. This is especially true in a company environment where you may have various projects, each using its own tools and practices, making it harder to move between projects.
	\item Use standard tools that are well known (e.g., Maven or Docker). That increases the likelihood that a newcomer would already know the tool and be able to develop for your project.
	\item It's good to have documentation about best practices but it's even better if the important ``must'' rules be enforced automatically by a checking tool (can be part of the build for example, or part of your IDE setup). For example instead of saying ``this @Unstable annotation should be removed after one development cycle'', you could write a Maven Enforcer rule (or a Checkstyle rule, or a Spoon rule) to break the build if it happens, with a message explaining the reason and what is to be done. Usually humans may prefer to have a tool telling them that they haven't been following the best practices documented at such location.
	\item Have a bot to help you discover documentation pages about a topic. For example, by having a chat bot located in the project's chat, that when asked about will give you the link to it.
	\item Projects must have a medium to ask questions and get fast answers (such as a chat tool). Forum or mailing lists are good but less interesting when onboarding when the newcomer has a lot of questions in the initial phase and requires a conversation.
	\item Have an answer strategy so that when someone asks a question, the doc is updated (new FAQ entry for example) so that the next person who comes can find the answer or be given the link to the doc.
	\item Mentoring (human aspect of onboarding): have a dedicated colleague to whom you're not afraid to ask questions and who is a referent to you.
	\item Supporting a variety of platforms for your software will make it simpler for newcomers to contribute to your project.
	\item Split your projects into smaller parts. While it's hard and a daunting experience to contribute to the core code of a project, if this project has a core as small as possible and the rest is made of plugins/extensions then it becomes simpler to start contributing to those extensions first.
	\item Have some interactive tutorial to learn about your software or about its development. A good example of nice tutorial can be found at \ahref{http://web.archive.org/web/20180328080916/https://www.katacoda.com/}{www.katacoda.com} (for example for Docker, \ahref{http://web.archive.org/web/20180608001913/https://www.katacoda.com/courses/docker}{https://www.katacoda.com/courses/docker}).
	\item Human aspect: have an environment that makes you feel welcome. Work and discuss how to best answer Pull Requests, how to communicate when someone joins the project, etc. Think of the newcomer as you would of a child: somebody who will occasionally stumble and need encouragement. Try to have as much empathy as possible.
	\item Make sure that people asking questions always get an answer quickly, perhaps by establishing a role on the team to ensure answers are provided.
	\item Last but not least, an interesting thought experiment to verify that you have some good onboarding processes: imagine that 1000 developers join your project / company on the same day. How do you handle this?
\end{itemize}

If you own a project, we would be interested to hear about your ideas and how you perform onboarding. You could also use the list above as a way to measure your level of onboarding for your project and find out how you could improve it further.


\abstracttitle{DevOps and the Need for Improved Evidence from Industrial Studies}
\label{sec:evidence_industrial_studies}
\abstractauthor[Lucy Ellen Lwakatare and Sigrid Eldh]{Lucy Ellen Lwakatare (University of Oulu, FI) and \\Sigrid Eldh (Ericsson AB - Stockholm, SE)}
\license

\subsubsection{Challenge}

One challenge, particularly when considering DevOps transformation in organizations, is the limited availability of supporting evidence of its implementation and quantifiable value in companies. 
In academic forums, anecdotal evidence is often presented in form of experience reports from leading innovative companies like Facebook and Netflix. For other software intensive organizations, particularly of safety critical, public sector and embedded domain, there exists very limited scientific support of DevOps implementation. 
This raises the question on whether the different facets of DevOps implementation and the corresponding impacts from pioneering companies are generalizable considering various constraints, e.g., legal and context, that do not necessarily make DevOps transformation easier. 
Furthermore, it is not clear what measures/indicators are established to clearly assess the value achieved through DevOps transformation. 
This is important because it is severally reported that DevOps adoption in a software-intensive organization often brings about changes to existing organizational structures, system architecture, e.g., to microservices and software development process, e.g., incorporation of deployment pipeline.

While emerging scientific studies make an initial step to describe the phenomenon, including to define and describe DevOps concept, practices and lessons learned from its early adoption, additional empirical studies are needed to crystallize core contribution of DevOps phenomenon. 
Practices should (at best) be explained and categorized before and after the DevOps introduction, where particular note should be to evaluate ``loss'' of practices (1) and ``new enabled practices'' not previously practiced (2). 
One should also take into account contextual factors that impact the success or failures.

\subsubsection{Our Proposal}

Based on this need, we propose studies that focus on identifying
\begin{itemize}
	\item What are software-intensive companies that have adopted DevOps doing, i.e., \textit{what aspect of DevOps practice is taken into use, in what context}
	\item What is (1) lost, (2) modified, (3) unchanged, and (4) added in existing practices as a result of DevOps in context. The numbers 1--4 are presented in Figure~\ref{fig:process_practices} below.
	\begin{itemize}
		\item Building evidence for what aspects of practices lost is not making things better - and how to make sure these good practices are sustained 
	\end{itemize}
	\item The added value given by DevOps and to which stakeholder(s)
	\begin{itemize}
		\item Giving new possibilities that did not exist before
	\end{itemize}
	\item Measures/indicators used to assess the value, and at what duration/time period as well as whether they can be monetized (\$)
	\item More support for change and transformation -- how to address issues in process change ``Onboarding''
\end{itemize}

\begin{figure}[h]
	\centering
	\includegraphics[width=0.5\columnwidth]{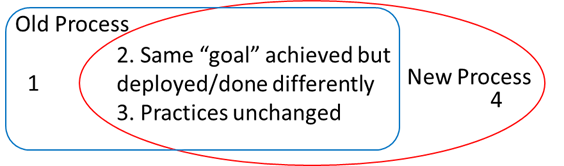}
	\caption{Comparing process practices in DevOps}
	\label{fig:process_practices}
\end{figure}

%


\section{Working Groups - Teaching}
\label{sec:teaching}

Starting with the talk of Christopher Parnin (see Section~\ref{sec:talk_parnin}) we shifted the focus of our discussion towards teaching DevOps-related concepts.
Three breakout groups formed tackling the topic from multiple perspectives involving the infrastructural challenges for hosting such a course, but also what concepts and technologies should actually be part of a curriculum. 

\abstracttitle{Designing a DevOps Course}
\label{sec:designing_devops_course}
\abstractauthor[Andrew Neitsch, Georgios Gousios, and Moritz Beller]{Andrew Neitsch (Cisco Systems Canada Co. - Toronto, CA), Georgios Gousios (TU Delft, NL), and Moritz Beller (TU Delft, NL)}
\license

\subsubsection{Introduction}

\begin{figure}[h]
	\centering
	\includegraphics[width=0.95\columnwidth]{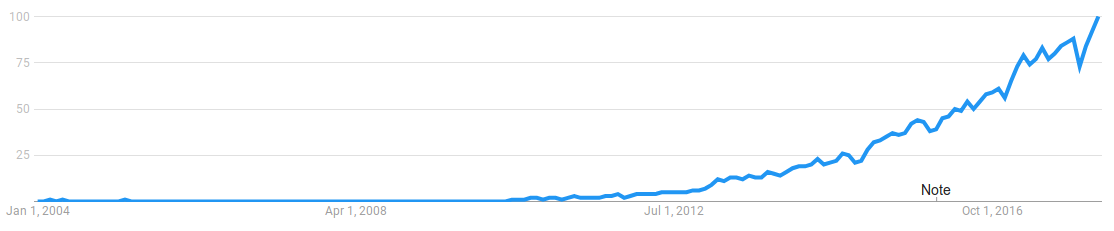}
	\caption{Google Trends for ``DevOps''. We can see the term ``DevOps'' pickup traction in 2012.}
	\label{fig:google_trends}
\end{figure}

Teaching a novel topic like DevOps (see Figure~\ref{fig:google_trends}), which is driven by industry needs, has always been a challenge for academia. 
It is not like teaching, say, an introductory course in compilers, where the basic theory has been codified in textbooks for decades, and the major difference between tools is a choice of target programming languages.

In this blog post, we discuss whether DevOps courses at university level should focus on teaching the principles or a set of tools, and if so, which principles, and which tools, and to which level of depth?

\subsubsection{Challenges in teaching DevOps}
	
The particular issues of infrastructure and operation automation that DevOps tries to solve only arise in fairly large industrial systems that have hit the boundaries of what can be achieved with manual human work. 
Naturally, most students are alien to this level of scale and therefore lack the understanding for why DevOps practices are needed. 
One way to make students see value in DevOps, could be to bring in the industrial perspective early-on in a course as a motivating example. Moreover, students usually appreciate guest lectures from industry. 
A firsthand story about a problem in the field, such as a website going down without anyone noticing until a major customer complained, can provide motivation for why these topics are important.

Another related challenge is that, while we seem to converge on a mutually shared understanding of DevOps, its boundaries are not yet clearly defined. 
As an evolving concept, it is not clear which material a good course on DevOps should cover. 
For example, to what extent should it cover testing on the Continuous Integration server? 
This is a choice left to the individual teacher and the surrounding courses offered at that university. 
Due to DevOps's contemporary nature, by the time we cover material in a course, it might be outdated.

Lastly, university courses usually aim at teaching principles rather than tool specifics. 
However, in DevOps, the distinction between what is tool-specific and what is general enough to serve as a principle is somewhat blurry. 
In some sense, the principles behind DevOps are easy to understand, but exactly their implementation in practice is hard. 
A large part of the complexity in DevOps stems from making specific tools interact successfully in a pipeline.

\subsubsection{Suggested learning objectives for developing new courses in DevOps}

\begin{figure}[h]
	\centering
	\includegraphics[width=0.80\columnwidth]{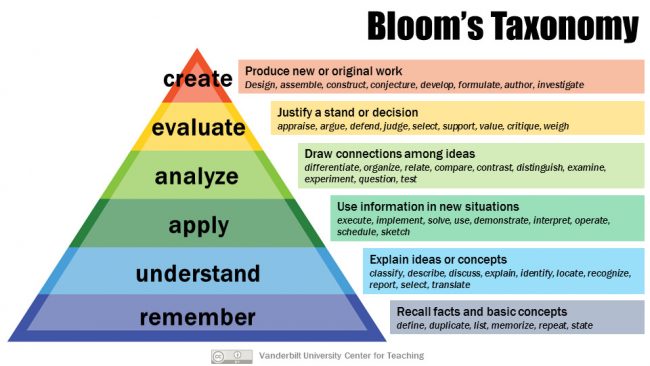}
	\caption{Bloom's Taxonomy of Learning Domains.}
	\label{fig:bloom_taxonomy}
\end{figure}

Courses are usually described in terms of learning objectives. 
What should students know after taking the course? 
There is not enough time to teach everything to a point of mastery, but Bloom's taxonomy\footnote{\ahref{http://web.archive.org/web/20180608002348/https://cft.vanderbilt.edu/guides-sub-pages/blooms-taxonomy/}{https://cft.vanderbilt.edu/guides-sub-pages/blooms-taxonomy/}} in Figure~\ref{fig:bloom_taxonomy} is useful to describe to what extent the objectives should be learned. 
It is a continuum from basic to very advanced knowledge.

\begin{figure}[h]
	\centering
	\includegraphics[width=0.85\columnwidth]{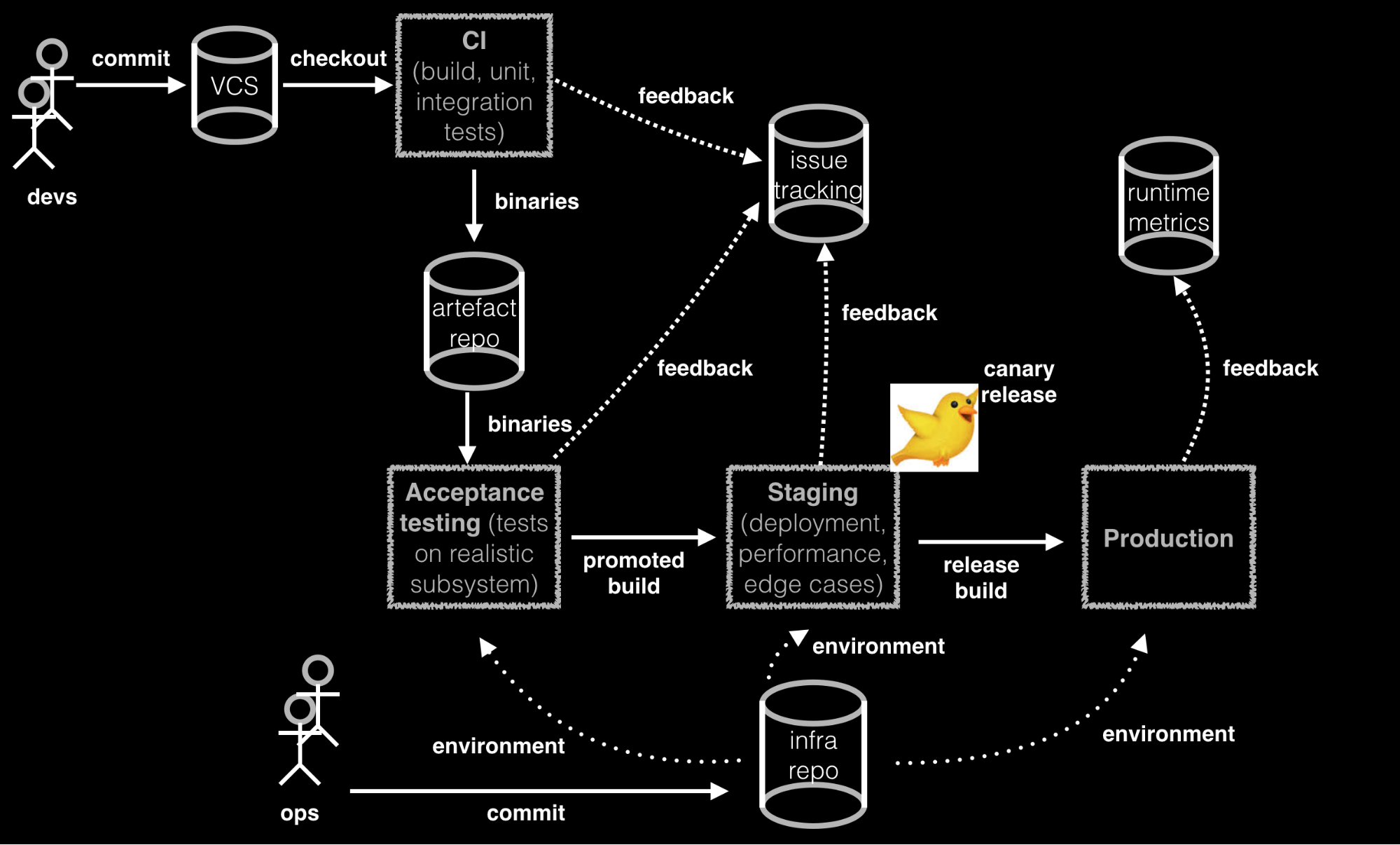}
	\caption{DevOps Workflow (source: Georgios Gousios).}
	\label{fig:devops_workflow}
\end{figure}

A good course on DevOps should prepare students with both the theory and the practice of continuous delivery pipelines and automated infrastructures. 
In that sense, it needs to emphasize both topics related to automating software engineering tasks and topics that have to do with automating the infrastructure. 
The description of our course relates to the DevOps pipeline view in Figure~\ref{fig:devops_workflow}.

In our discussions, we came up with the following topics that we believe are essential for students to understand. 
Initially, all students need to be up to speed with modern \emph{version control}; even though this is usually taught at basic software engineering courses in most modern programs, its significance in DevOps pipelines (everything starts with a commit!) makes it an important topic to revisit. 
Then, students also need to come up to speed with practices pertaining to \emph{automated testing}, especially in the context of using testing for quality assurance. 

\emph{Automated testing} can come in multiple flavors (e.g., unit testing, acceptance testing etc) that are being used in various stages in a DevOps pipeline; the students should learn to analyze what testing results mean and what their effects are on the actual product. 
\emph{Continuous integration} is usually the process that drives the whole pipeline; consequently the students must know how to apply appropriate tools, how to combine them with version control tools (that act as the continuous delivery trigger), how to apply appropriate static analysis tools, and how to store build artifacts (one important concept is immutability). 

\emph{Automated deployment} is the process of applying a build artifact in a production or testing environment; there is a big variation in the tools that can be used for this purpose and the process is also context specific, so we believe that students should learn to apply basic tools (e.g., scripting languages) before trying for specialized ones. 
The industry does seem to be on a path to convergence on using Docker and Kubernetes in the future, but traditional application deployments will still be around for a long time.

The above would make for good knowledge on continuous integration pipelines; what is missing is how to automate infrastructure to deploy on. 
For this, the students need a different set of, mostly practical, skills that usually is not part of computer science curricula. 
This means that a DevOps course needs to emphasize practical system administration and also how to automate it. 
Again, a plethora of tools, each with a different philosophy behind it, makes it difficult to extract and teach generic principles; thus, the students need to learn to apply one of them (e.g., Ansible or Puppet) in practice.

\emph{Production monitoring} is to ensure that, once deployed, a service stays up, and what to do when it does not. 
There are various third-party services that can email or text someone when something goes down. 
More detailed metrics gathering can be used for capacity planning, and can provide useful business metrics such as how many customers there are and how they are using the application.

\subsubsection{Recommendations}

In the following, we identified six distinct concepts that a DevOps course should cover. 
We annotate each concept with the depth of knowledge level that we would expect to be taught at a university-level course. 
We think that in principle the concepts covered in a DevOps course can be understood by Bachelor-level students. 
However, practical constraints might make it feasible to only teach a DevOps course at the Master level, even though some concepts such as CI might be covered earlier in the Bachelor. 
In the last column we give an example of tools that should be covered for the given concept.

\begin{table}[h!]
	\centering
	\begin{tabular}{p{0.27\columnwidth}ll}
		\toprule
		\textbf{Concept} & \textbf{Bloom Knowledge Level} & \textbf{Example Tools} \\ 
		\midrule
		Version Control & Apply & Git, GitHub \\
		Automated tests: unit tests, integration testing, acceptance testing, ... & Analyze & xUnit, Selenium \\
		Continuous Integration & Apply & Travis CI, Jenkins \\
		Automated Deployments & Apply & Shell scripts, Docker, Kubernetes \\
		Automated Infrastructure & Apply & Chef, Puppet, Ansible, SaltStack \\
		Production Monitoring & Evaluate & Pingdom, Grafana, ELK Stack \\
		\bottomrule 
	\end{tabular} 
	\caption{Concepts of a DevOps course mapped to Bloom Knowledge Levels}
	\label{tab:designing_devops_course}
\end{table}


\abstracttitle{What are the Infrastructural Challenges in Hosting a DevOps Course?}
\label{sec:infrastructural_challenges}
\abstractauthor[Weiyi (Ian) Shang, Shane McIntosh, Christopher J. Parnin, Sarah Nadi, Benjamin Danglot, Zhen Ming (Jack) Jiang, Lucy Ellen Lwakatare, Hui Song]{Weiyi (Ian) Shang (Concordia University - Montreal, CA), Shane McIntosh (McGill - Montreal, CA), Christopher J. Parnin (North Carolina State University - Raleigh, USA), Sarah Nadi (University of Alberta - Edmonton, CA), Benjamin Danglot (INRIA Lille, FR), 
Zhen Ming (Jack) Jiang (York University - Toronto, CA), Lucy Ellen Lwakatare (University of Oulo, FI), Hui Song (SINTEF ICT - Oslo, NO), and Oscar Luis Vera P\'{e}rez (INRIA Rennes, FR)}
\license

Due to the wide adoption of DevOps in practice, there is a clear need to introduce such DevOps practices into the higher education curriculum, as a course, or even throughout the educational program. 
While clearly there is benefit in offering such knowledge in the school, the course(s) comes with a cost. 
The question that we would like to touch on in this blog post is simply, what are the challenges that we face, when offering a DevOps course.

\subsubsection{Context of the DevOps course}

DevOps is a very practical discipline which has received a lot of attention from industry and academia. 
The designed course will consists of lectures, hands-on labs, and projects. 
Hence, in order to provide practical hands-on experience with the students, we intend to provide accessible infrastructures representing state-of-the-art practices for students to experiment with. 

\subsubsection{Hardware (virtual hardware) support}

\paragraph*{Do-it-yourself or using a service?}

Free (sort of) services and provisioning tools are available to support in each phase of the pipeline. 
For example, Docker and Vagrant can help in provision a service of particular phases in the DevOps pipeline. 
Travis CI or CircleCI can be used as readily available services for the build and testing phase of the pipeline, while Jenkins servers can be setup to assist in the build and testing phase as well. 
Finally, Heroku can be used as a deployment service.
There is a decision to make on whether a group of student should take off-the-shelf services to accomplish each phase or pick a readily available service to help.
In general, the decisions can be made by considering two aspects: 
\begin{enumerate}
	\item The focus of the course project. While the service facilitates having the pipeline ready, the students may miss the opportunities to exercise or learn what is under the hood. For example, using Travis may miss the chance to learn how to provision and install a build/test server (e.g., Jenkins). If such exercise is considered important by the instructor of the course, those readily available services may be avoided. 
	\item The ability on special needs. A readily available service may on one hand accomplish many tasks hassle free, while it also may lack the ability to be customized. For example, an off-the-shelf hosting solution like Heroku may not have the flexibility to implement complex deployment strategies (e.g., feature toggle consolidation, blue/green deployment).
\end{enumerate}

The challenge does not stop here. 
Even without a readily available service, there is still a level of detail you want the student to experience. 
For example, a student may want to use a Docker image to host a Jenkins server, which may let the student miss the chance of experiencing deploying a Jenkins server from scratch. 

\paragraph*{In-house or on the Cloud?}

If the students are required to provision their own service to support a phase in the pipeline, they need to infrastructure support to host such services. 
Here are the availabilities:
\begin{enumerate}
	\item \emph{Public provisioning providers:} AWS, Azure, Digital Ocean, etc. You name it. The list goes on. But using these providers is not free. You will be noticed if your student receive a thousand dollar bill. In fact, we know that there are educational funding supports from some providers (e.g., AWS and Azure). However, it should not be considered as a long term plan. Another issue with leveraging public provisioning providers is the issue of privacy. The course material may be considered as an intellectual property of the instructor and/or the university. You may risk the chance of breaking the rule if the material is shared or even stored by public provisioning providers.
	\item \emph{Institutional and regional provider:} Universities, or governments often host their own infrastructure that can be used to host services for educational needs. Since those providers may not be optimized for the course context, local services may be needed to support the pipeline. For example, a local maven repository within the provider's local network may need to be created.
	\item \emph{Students' local machines:} The last resort of these infrastructure is to provision on student's machines, e.g., their laptops by creating VMs or Docker images. An apparent issue of such an option is the ability of host multiple VMs or images on their laptops. More importantly, the solution seems to be easier to try out but may be different to running with providers, which is more realistic in practice. Moreover, the student do not get the chance to exercise the operation of the system in the field.
\end{enumerate}

\subsubsection{Material (or process) support}

\paragraph*{How can students be graded?}

First off, the form of examinations are not preferred for such a practical course. 
The students should be graded based on their project of engineering a DevOps pipeline. 
Here are the options that are available: 
\begin{itemize}
	\item Screencast,
	\item Coverage graphs, 
	\item Analysis results, 
	\item (or any empirical evidence that they have done the work)
\end{itemize}

However, grading such materials can be strongly subjectively biased, and on the other hand, resource costing. 

\paragraph*{How to include development history through the pipeline during the course?}

Although the focus on the course is the DevOps pipeline, there is no pipeline useful without actual development on the subject system code. Requesting students to make development to the actual subject system code becomes a simple but naive solution since 1) the student may already be overwhelmed by the pipeline itself, and 2) more importantly, it can bring confusion about what is the real focus of the course. 
Two possible directions are clear at this stage:

\textbf{Fake the development:}
Since the development is not the real focus of the course, it does not seem to be of much value to do a real development on the side. 
With this spirit, simulated development, (as simple as randomly change a line of code), can be done to exercise the pipeline. 

\textbf{``Outsourcing'' the development:}
No we do not mean to hire people to develop a subject project on the side. 
In reality, students often need to through team projects in various courses, such as software design, software process, etc. 
With those courses typically including a development project, the DevOps course can be coordinated with those courses to provide the pipeline of those projects, while using the development of those projects to exercise the pipeline.
In such cases, coordination issues becomes the showstopper. 
Here are some (while we believe more exist):
\begin{itemize}
	\item Not every professor wants to include DevOps.
	\item Not every student has the same prerequisites or having pair classes
	\item The different courses needs to sync their teaching schedule.
	\item (the list goes on)
\end{itemize}

\subsubsection{Summary}
To conclude, hosting a DevOps course is not a trivial task. 
Due to the nature of the subject, it comes with both technological and non-technological challenges. 
On the one hand, we encourage people start considering such a subject in their courses, while proper solutions to the above mentioned challenges may be considered to deliver a successful course. 


\abstracttitle{Teaching Topics Requirements}
\abstractauthor[Bram Adams, Tamara Dumic, Foutse Khomh, and Andy Zaidman]{Bram Adams (Polytechnique Montreal, CA), Tamara Dumic (Ericsson Research - Stockholm, SE), Foutse Khomh (Polytechnique Montreal, CA), and Andy Zaidman (TU Delft, NL)}
\license

Three challenges that evolved from our working group about typical requirements for setting up and conducting a DevOps courses will be discussed in the following.
We conclude with a brief discussion on topics to consider for a DevOps course and the final question covers potential ways to assess students in such a course.

\subsubsection{Challenge 1: DevOps has a strong dependency on other courses}

We feel that DevOps is a cross-cutting topic. 
There are elements of software testing in it, but also relations to software architecture, software design, software requirements and software process.

For example, a microservice architecture makes it easy to frequently redeploy specific services, without having to take the entire system offline. But this requires advance thinking at the level of the software architecture. 
Also the design level can be important, as at this level the observability of the state of the system can be influenced, which in turn influences how well you can test your system.

Similarly, the requirements can stipulate that rapid releasing is necessary, or that advanced monitoring of the deployed software is necessary. 
Both can be traced back to DevOps principles. 
So this stage can also influence the decision making with regards to DevOps.

Your software process (e.g., waterfall, agile, ...) is also an influencing factor on whether and how you want to install your DevOps pipeline.

Our idea is that it would be good that already in these courses this link is highlighted, but vice versa, if a separate course on release engineering is set up, this course should also reflect on the importance of these other subfields, as they are influencing the scope of DevOps to a large degree.

Also, many software engineering programs have project courses in which students develop and release software, such courses could also serve as a vehicle to distill DevOps practices to student throughout their education.

\subsubsection{Challenge 2: Raise the awareness of students of (1) the impact their commits have on the whole project and (2) the scale of the release engineering process} 

Students typically think that writing code and taking care of merge conflicts when pushing to version control is all that needs to be done. 
However, the reality is that students have a hard time imagining what the potential impact is of making changes to their component when they are working in a much larger project that has many build dependencies. 
By installing a DevOps pipeline for their school projects their insight will increase, but even then this does not compare to some massive and complex build processes that can be seen in large companies.

We find that this is an argument to also involve guest speakers from industry, as they are likely able to show the scale issues that arise.

\subsubsection{Challenge 3: How to keep the course relevant amidst rapid changes in concepts and technologies}

As the field of DevOps is still emerging and people are finding their way in the area, it is still not entirely clear which set of technologies will emerge as ``standard''. 
While it seems that Ansible and Docker are gaining traction here, it might be that new technologies or rival technologies will still emerge. 
Early experiences are also that existing tools are rapidly evolving still, thus requiring frequent updates to course material.

An important question that we have asked ourselves during the discussion is whether we should focus our teaching on technology or on principles. 
Or whether the principles can be taught through technology. 
We feel that the technology should be taught and that this should be enough for most students to get the necessary awareness and insights into the general principles.

\subsubsection{Topics to cover in a DevOps course}

Given the cross-cutting nature of DevOps activities, there is a wide range of skills that students need to acquire to able to contribute efficiently to a software release. 
Hence, as mentioned earlier it is important that courses that cover key activities of the software development and release process includes aspects of DevOps. 
However, setting up and running an efficient release pipeline requires specific knowledge that should be covered in a DevOps course. 
Students in a DevOps course should be taught configuration management with an artifact repository, i.e., how to manage dependencies, traceability between software products and requirements, etc. 
Another important topic is dynamic scheduling/provisioning of test environments across different levels (``what hardware and software do I need without blocking any limited resource unnecessarily?''). 
We should also teach students to build and provision efficient release pipelines (i.e., ensuring that their pipelines are capable of catching issues efficiently without making an excessive use of the limited resources available) and to ensure the security of these pipelines to prevent malwares from slipping into released products.  
Other interesting topics to cover are available in the summary of the 2016 workshop on DevOps education\footnote{\url{https://drive.google.com/open?id=1LwNLmF252uFtQPp6Q6QEmP0am-OBul_n}}. 
The Linux Foundation also provides interesting pointers on topics to cover in a DevOps course. 

\subsubsection{How to assess students in a DevOps course?}

To assess the knowledge acquired by students we propose the following mechanisms: 
\begin{itemize}
	\item Cross-evaluation of assignments/projects 
	\item Make students write some kind of reflection about the DevOps experience of the class (a 2 page paper)
	\item Make the students do screencasts to explain their work
	\item Write a personal diary of an operator after a hard day at work
\end{itemize}

Pre-requisite skills include networking, virtualization, and troubleshooting skills.
Further, different kinds of testing could be involved and assessed: unit, functional, non-functional testing, staging, etc.
Code could be assessed as well, e.g., clean code: design patterns, principles, best practices.
Further, architectural styles (e.g., microservices) applied, aggregated pipelines, and quality assurance in general.

\begin{itemize}
	\item Requirement courses, architecture courses, quality assurance course should cover ``Ops'' issues

	\item Teaching DevOps principle, environments, release pipelines

	\item Hands on experience with some specific tools for release through a project $ \rightarrow $ it is better for students to have experience with some specific tools than no experience at all and only broad knowledge of concepts

	\item Huge application: system-level tests are too complicated for individual developers (too many interactions) $ \rightarrow $ how to obtain feedback that is concrete enough for developers to know how to fix the issue?

	\item Industry/practice: missing knowledge on different levels of testing, link with clean code, mock testing, 10s of pipelines: follow your commit until certain level

\end{itemize}


Not necessarily directly linked to assessing students, but teaching support is also an important aspect to take into account.
Thus, how to train teaching assistants for such a DevOps course?
One idea would be to ask the best students of the previous year to TA the course.
Or, to limit the first edition to a few students and manage the course and then in the subsequent editions of the course, pick the best students of the previous edition to TA.



\section{Panel Discussions}
\label{sec:panel}

After the talks from Sigrid (see Section~\ref{sec:talk_eldh}) and Hyrum (see Section~\ref{sec:talk_wright}), we discussed two main topics, namely the architecture(s) that enable DevOps (Section~\ref{sec:panel_architecture}) and what to do with the data collected in the Ops phase (Section~\ref{sec:panel_data}). 
Finally, we concluded the seminar with discussing potential next steps.

\abstracttitle{Microservices for DevOps}
\label{sec:panel_architecture}
\abstractauthor[All Participants]{All Participants}
\license

The architecture discussion that followed the talks went pretty quickly towards microservices. 
While there was consensus that a microservice architecture is not the only possible architecture for enabling DevOps, it seems that microservices are associated with the concept of DevOps quite frequently. 
The fact that these services are relatively small and loosely coupled, make it easy for them to get re-deployed while the system is running. Nevertheless, also more monolithic architectures can be build and deployed in a DevOps fashion, just consider mobile apps for example that might also send in post-deployment data.

Another aspect of the discussion is then how we test this microservice architecture. 
Each microservice can be tested at design time, but how the microservices interact can only be tested at runtime through some form of online testing, thus collecting data on how these microservices interact becomes all the more important.
Related to the previous point, the discussion also went into the direction that some companies actually need to reverse engineer the interactions between the microservices to get an understanding of the actual architecture of the microservices system. 
So some of the business logic is in how the services interact, and not only in the microservices themselves anymore. 
This again raises the importance of online testing.


\abstracttitle{Dealing with Large Amounts of Data}
\label{sec:panel_data}
\abstractauthor[All Participants]{All Participants}
\license

DevOps processes generate huge amounts of data, which need to be acted upon in order to feed back operations insights to development teams (in the broad sense). 
Needless to say, those insights can be pretty well hidden within heaps of noise.

For example, the popular TravisTorrent data set providing 60 attributes of 3.7 million Travis CI build jobs is based on a raw data set of over 2 TB of build logs (\ahref{http://web.archive.org/web/20180608002525/https://blog.travis-ci.com/2017-01-16-travis-ci-mining-challenge/}{https://blog.travis-ci.com/2017-01-16-travis-ci-mining-challenge/}). 
To make analysis of this data feasible for researchers, the logs were aggregated into 3.2 GB of SQL dumps, focusing on the most commonly requested attributes of the builds.

In industry, the scale of things is even more extreme due to the wide variety of telemetry data gathered for each transaction and event. 
In 2016, Netflix' Keystone data pipeline\footnote{\ahref{http://web.archive.org/web/20180608002525/https://medium.com/netflix-techblog/evolution-of-the-netflix-data-pipeline-da246ca36905}{https://medium.com/netflix-techblog/evolution-of-the-netflix-data-pipeline-da246ca36905}} had to process 500 billion events (1.3 petabyte of data!) per day. 
At peak times, 8 million events (24 GB of data) are generated per second(!). 
In 2017, Twitter's event logs\footnote{\url{https://events.static.linuxfound.org/sites/events/files/slides/Routing\%20Trillion\%20Events\%20per\%20day\%20\%40Twitter.pdf}} processed more than a trillion of events per day, amounting to 3 PB of uncompressed data per day.

These numbers are increasing each year, forcing companies to also scale up and improve their data pipeline. 
In December 2015, Netflix' Keystone pipeline was already the third incarnation of the company's data pipeline. 
Apart from being able to process such data in real-time, companies need to consider the need for long-term storage of this data, which adds additional nightmares. 
Legislation like the Sarbanes-Oxley Act\footnote{\ahref{http://web.archive.org/web/20180608002525/https://en.wikipedia.org/wiki/Sarbanes}{https://en.wikipedia.org/wiki/Sarbanes}\%E2\%80\%93Oxley\_Act} requires certain kinds of data to be retained for a long time, with certain requirements regarding confidentiality and privacy.

\noindent Given these facts, this grand challenge focuses on the following major questions:
\begin{itemize}
	\item How to identify the types of operations data that are useful in the context of DevOps?
	\item How to efficiently collect the identified data?
	\item How to efficiently process the identified data in order to extract actionable insights for ``Dev''?
	\item How to back the data up in an efficient way, safeguarding privacy and confidentiality?
	\item How to give data scientists access to portions of the data for advanced analysis, without exposing too many privacy details?
	\item How to succinctly report the data back to ``Dev''?
\end{itemize}


\abstracttitle{Next Steps}
\label{sec:panel_next}
\abstractauthor[All Participants]{All Participants}
\license

One idea raised was to launch a follow-up event of the workshop on Release Engineering at ICSE'19. 
We briefly discussed similar events that are upcoming or took place recently:
\begin{itemize}
	\item DevOps 2018: Bertrand Meyer organized the ``First international workshop on software engineering aspects
	of continuous development and new paradigms of software production and deployment''\footnote{\ahref{http://web.archive.org/web/20180608002526/https://www.laser-foundation.org/devops/2018/}{https://www.laser-foundation.org/devops/2018/}}
	\item Philipp Leitner was involved in organizing the ``Vienna Software Seminar''\footnote{\ahref{http://web.archive.org/web/20180608002528/https://vss.swa.univie.ac.at/2017/}{https://vss.swa.univie.ac.at/2017/}} covering Continuous Delivery from an architecture perspective.
\end{itemize}

A potential reboot of the ICSE workshop could focus on technical challenges.
Furthermore, it should be no problem if the workshop occurs only once, twice or three times.
``Let us not fear to stop a workshop series.''
Another idea raised was to take a look at industry conferences such as DEVOXX\footnote{\url{https://devoxx.com/}}, Velocity\footnote{\ahref{http://web.archive.org/web/20180608002528/https://conferences.oreilly.com/velocity}{https://conferences.oreilly.com/velocity}}, or FOSDEM\footnote{\ahref{http://web.archive.org/web/20180608002530/https://fosdem.org/2018/}{https://fosdem.org/2018/}}. 
A Dagstuhl representative also mentioned that Dagstuhl can also host summer schools.

~\\
\noindent We also discussed potential goals of such a follow-up event:
\begin{itemize}
	\item Education
	\item Community building
	\item Involve academia, software industry who contribute to DevOps, software industry who want to adopt DevOps, clients who see their software suppliers adopt DevOps
	\item Crystalize the core research problems in DevOps
	\item Awareness of the human factors
\end{itemize}

\noindent A further item of our discussion was on how we can reach out to ``Ops'' people:
\begin{itemize}
\item Who are they? What do they study? What conferences do they attend? (e.g., Velocity, AWS conference)
\item They build tools that enable developers to do their job better (DB, systems, networks)
\item From an academic perspective, they are in the ``network and system'' department
\item They are people who like to see a system running
\item LISA\footnote{\ahref{http://web.archive.org/web/20180608002531/https://www.usenix.org/conference/lisa18}{https://www.usenix.org/conference/lisa18}} is a Usenix conference where Ops people go
\item Related to DB, systems, resource management
\end{itemize}

One challenge that we did not touch upon is how to adopt DevOps for embedded systems. 
The key issue there is that the gap between Dev and Ops is very difficult to bridge. 
That is mainly because the deployment on the embedded platforms is very specific, which makes it hard to automate.

\end{document}